\documentclass[%
 aip,
 jmp,
 amsmath,amssymb,
 reprint,%
]{revtex4-1}

\usepackage{graphicx}
\usepackage{dcolumn}
\usepackage{bm}

\usepackage[utf8]{inputenc}
\usepackage[T1]{fontenc}
\usepackage{mathptmx}
\usepackage{etoolbox}

\makeatletter
\def\@email#1#2{%
 \endgroup
 \patchcmd{\titleblock@produce}
  {\frontmatter@RRAPformat}
  {\frontmatter@RRAPformat{\produce@RRAP{*#1\href{mailto:#2}{#2}}}\frontmatter@RRAPformat}
  {}{}
}%
\makeatother
\begin{document}

\preprint{AIP/123-QED}

\title{Study of stationary rigidly rotating anisotropic cylindrical fluids with new exact interior solutions of GR. 2. More about axial pressure.} 



\author{M.-N. C\'el\'erier}
\email{marie-noelle.celerier@obspm.fr}
\affiliation{Laboratoire Univers et Th\'eories, Observatoire de Paris, Universit\'e PSL, Universit\'e Paris Cit\'e, CNRS, F-92190 Meudon, France}


\date{\today}

\begin{abstract}
This article is the second in a series devoted to the study of spacetimes sourced by a stationary cylinder of fluid rigidly rotating around its symmetry axis and exhibiting an anisotropic pressure by using new exact interior solutions of General Relativity. The configurations have been specialized to three different cases where the pressure is on turn directed alongside each principal stress. The two first articles in the series display the analysis of the axial pressure case. Indeed, the first axial class published in Paper 1 is merely a special case. It is recalled here and its properties are revised and supplemented. Moreover, a fully general method aiming at constructing different classes of such solutions is displayed. This method described in the present paper, Paper 2, represents a key result of this work. It is exemplified and applied to two new classes of solutions depending on a single constant parameter. One of them, denoted Class A, is shown to verify every conditions needing to be satisfied by a fully achieved set of exact solutions: axisymmetry and, when appropriate, regularity conditions, matching to an exterior vacuum, proper metric signature, weak and strong energy conditions. Other properties and general rules are exhibited, some shedding light on rather longstanding issues. Astrophysical and physical applications are suggested.
\end{abstract}


\maketitle 

\section{Introduction} \label{intro}

To study fundamental issues in General Relativity (GR) some symmetries of spacetime are often used to somehow alleviate the complication of Einstein's field equations. Since it is hard to work with merely one Killing vector, two Killing vectors can be considered, leading, e. g., to cylindrical symmetry. Hence, rotating cylindrically symmetric spacetimes have been extensively investigated for a number of different purposes \cite{G09,S09,B20}. A further simplification, first performed by Lewis for vacuum spacetimes \cite{L32}, can be implemented by adding a third Killing vector implying stationarity.

However, a gravitational source composed of a rotating cylinder of matter, infinitely extended along its axis of symmetry and generating a spacetime deprived of asymptotic flatness, can hardly be taken at face value to represent exactly a rotating astrophysical object made of standard matter. But this is not the final point since exact solutions have mainly been used, in astrophysics or in cosmology, to approximate some kind of physical processes or to roughly model some particular region of spacetime. Hence any quasi-cylindrical object, such as a gravitational source composed of a rotating fluid issuing, e. g., jets along its axis of rotation such that its axial extension is huge as regards the length of its radius might be validly approximate by such solutions. Another particularly promising field for their application is the network of cosmic superstrings developed in the framework of string/M-Theory, which formed before inflation and stretched up to macroscopic length scales afterwards \cite{C10}. Moreover, the study of such spacetimes yields both interesting mathematical and physical results.

The present article is the second (referred to as Paper 2) in a set devoted to the study of the gravitational influence of anisotropic pressure on interior spacetimes sourced by stationary rotating cylindrically symmetric fluids with the use of exact solutions of the Einstein equations. Exact solutions to the field equations of GR for interior spacetimes are known to be much more difficult to identify than their vacuum counterparts. Moreover, even though such solutions implying dust or a perfect fluid as their source have been exhibited in the past \cite{G09,S09,K75}, no exact rotating cylindrical solutions involving an anisotropic pressure were currently known before the publication of the first article in the present series,

In this companion paper, referred to as Paper 1 \cite{C21}, a particular class of solutions describing the interior of a cylindrically symmetric rigidly rotating stationary fluid with axially directed pressure has been displayed and analyzed. This particular class of spacetimes can be considered as a first step towards a more general study of anisotropic pressure in cylindrical symmetry. Indeed, a series of three different anisotropic types of interior spacetimes has been found. Each type includes a number of classes of interior  spacetimes gravitationally sourced by a stationary rigidly rotating fluid whose principal stresses are vanishing in pair.

In present Paper 2, general properties of exact solutions to the field equations are displayed for the case of axially directed pressure. Since the corresponding problem exhibits one extra degree of freedom, a general method for integrating different classes of such solutions is developed. This method is then applied to a new simple example, which is thoroughly integrated and analysed both from a mathematical and a physical point of view. Another fully integrated solution is displayed and analysed in Appendix B such as to provide an example showing how a mathematically and physically robust solution can be apparently ruled out on the ground of not fulfilling the ''regularity condition'' while this condition (i) is not necessary since the metric functions do not diverge on the axis, (ii) is therefore redundant and imposes unnecessary constraints leading to a bad behaviour of the metric.

The two other types, i.e., for azimuthally directed and radially directed pressure, are presented in two companion papers, denoted Paper 3 and Paper 4 respectively. For further astrophysical purpose, all these spacetimes, those of the present Paper 2 included, are matched to an exterior Lewis-Weyl vacuum. Actually, the Lewis solution describes a vacuum spacetime gravitationnaly sourced by a ''matter'' cylinder in stationary rotation around its symmetry axis \cite{L32}. The Weyl class includes the solutions whose constant parameters appearing into the metric functions are real numbers as well as the parameters appearing in the present new interior solutions. They are also required to verify the axisymmetry condition and their behaviour as regards regularity on the axis is discussed. 

The present paper is organized as follows: in Sec. \ref{sis}, the stationary cylindrically symmetric line element which will be used for the present purpose is set up and the field equations together with the corresponding Bianchi identity are displayed. In Sec. \ref{main}, the main equations to be solved are constructed from the field equations. In Sec. \ref{method}, the key result of this work is exhibited as a general method describing step by step the recipe for generating a number of different solutions to the axial pressure issue. Then the solution displayed in Paper 1 is recalled and completed in Sec. \ref{paper1}. Subsequently, the recipe of Sec. \ref{method} is exemplified in Sec. \ref{example} where a new fully integrated class of solutions constrained by a number of mathematical and physical requirements, i.e., axisymmetry, regularity and junction conditions, properties of the mathematical functions involved, metric signature, hydrodynamical constraints, is characterized and important mathematical and physical properties pertaining to these solutions are analyzed. Two main classes are identified. However, only that denoted Class A happens to verify the whole set of imposed constraints.  Section \ref{versus} is devoted to a comparison between the solutions described in Paper 1 and the new ones pertaining to Class A. The conclusions are displayed in Sec. \ref{concl}. Two Appendices are also provided. In Appendix A, a physical interpretation of the parameter $c$ generically appearing in the solutions of this series is proven. In Appendix B, another class of solutions to the field equations is displayed and analyzed in order to examine the influence of improperly imposing the ''regularity condition''.

\section{Equations determining the spacetime inside the source} \label{sis}

As in Paper 1, a rigidly rotating stationary cylindrically symmetric anisotropic nondissipative fluid bounded by a cylindrical surface $\Sigma$ whose principal stresses $P_r$, $P_z$ and $P_\phi$ obey the equation of state $P_r= P_{\phi}=0$ is considered here. With $\rho$, the energy density of the fluid, $V_\alpha$, its timelike 4-velocity, and $S_\alpha$, a spacelike 4-vector, satisfying
\begin{equation}
V^\alpha V_\alpha = -1, \quad  S^\alpha S_\alpha = 1, \quad V^\alpha S_\alpha = 0, \label{fourvec}
\end{equation}
the stress-energy tensor can be written as
\begin{equation}
T_{\alpha \beta} = \rho V_\alpha V_\beta + P_z S_\alpha S_\beta. \label{setens}
\end{equation}
To allow a proper matching to the exterior Lewis-Weyl metric, the $\partial_z$ Killing vector of the inside $\Sigma$ spacetime is assumed to be hypersurface orthogonal. The stationary cylindrically symmetric line element can therefore be written
\begin{equation}
\textrm{d}s^2=-f \textrm{d}t^2 + 2 k \textrm{d}t \textrm{d}\phi +\textrm{e}^\mu (\textrm{d}r^2 +\textrm{d}z^2) + l \textrm{d}\phi^2, \label{metric}
\end{equation}
where $f$, $k$, $\mu$ and $l$ are real functions of the radial coordinate $r$ only and the coordinates are conforming to the ranges
\begin{eqnarray}
-\infty &\leq& t \leq +\infty, \quad 0 \leq r\leq +\infty, \quad -\infty \leq z \leq +\infty, \nonumber \\ 
0 &\leq& \phi \leq 2 \pi, \label{ranges}
\end{eqnarray}
where the two limits of the $\phi$ coordinate are topologically identified. These coordinates are numbered $x^0=t$, $x^1=r$, $x^2=z$ and $x^3=\phi$.

In the case of rigid rotation a corotating frame can be chosen for the stationary source \cite{D06,CS20}. Thus, the 4-velocity of the fluid can be written as
\begin{equation}
V^\alpha = v \delta^\alpha_0, \label{r4velocity}
\end{equation}
$v$ being a function of $r$ only. The timelike condition for $V^\alpha$, provided by (\ref{fourvec}), reads therefore
\begin{equation}
fv^2 = 1. \label{timelike}
\end{equation}
The spacelike 4-vector $S^\alpha$ satisfying conditions (\ref{fourvec}) can be chosen as
\begin{equation}
S^\alpha = \textrm{e}^{-\mu/2}\delta^\alpha_2. \label{salpha}
\end{equation}
An auxiliary function $D$, first introduced by van Stockum \cite{vS37}, used to simplify the calculations but devoid of any physical meaning, is defined as
\begin{equation}
D^2 = fl + k^2. \label{D2}
\end{equation}

\subsection{Field equations} \label{fe}

Using (\ref{fourvec})-(\ref{D2}) into (\ref{setens}), one obtains the  components of the stress-energy tensor corresponding to the five non-vanishing components of the Einstein tensor. One can thus write the set of five field equations verified by the inside $\Sigma$ spacetime. With a new auxiliary function $h(r)$ defined as $h(r)\equiv P_z(r)/\rho(r)$, these equations, already displayed in Paper 1, are recalled here as
\begin{eqnarray}
G_{00} &=& \frac{\textrm{e}^{-\mu}}{2} \left[-f\mu'' - 2f\frac{D''}{D} + f'' - f'\frac{D'}{D} + \frac{3f(f'l' + k'^2)}{2D^2}\right] \nonumber \\
&=& \kappa\rho f, \label{G00}
\end{eqnarray}
\begin{eqnarray}
G_{03} &=&  \frac{\textrm{e}^{-\mu}}{2} \left[k\mu'' + 2 k \frac{D''}{D} -k'' + k'\frac{D'}{D} - \frac{3k(f'l' + k'^2)}{2D^2}\right]  \nonumber \\
&=& - \kappa\rho k, \label{G03}
\end{eqnarray}
\begin{eqnarray} 
G_{11} &=& \frac{\mu' D'}{2D} + \frac{f'l' + k'^2}{4D^2} = 0, \label{G11}
\end{eqnarray}
\begin{eqnarray}
G_{22} &=& \frac{D''}{D} -\frac{\mu' D'}{2D} - \frac{f'l' + k'^2}{4D^2} = \kappa \rho h \textrm{e}^\mu, \label{G22}
\end{eqnarray}
\begin{eqnarray}
G_{33} &=&  \frac{\textrm{e}^{-\mu}}{2} \left[l\mu'' + 2l\frac{D''}{D} - l'' + l'\frac{D'}{D} - \frac{3l(f'l' + k'^2)}{2D^2}\right] \nonumber \\
&=& \kappa \rho \frac{k^2}{f}, \label{G33}
\end{eqnarray}
where the primes stand for differentiation with respect to $r$.

\subsection{Bianchi identity}

This equation has also been disclosed in Paper 1. Recall however that writing the stress-energy tensor conservation is analogous to writing the Bianchi identity
\begin{equation}
T^\beta_{1;\beta} = 0. \label{Bianchi}
\end{equation}
The 4-velocity of the fluid $V^\alpha$ given by (\ref{r4velocity}) and the space-like vector $S^\alpha$ given by (\ref{salpha}) are inserted into the stress-energy tensor (\ref{setens}). Then, using (\ref{metric}) and (\ref{timelike}), the Bianchi identity (\ref{Bianchi}) reduces to
\begin{equation}
T^\beta_{1;\beta} = \frac{1}{2} \rho \frac{f'}{f} - \frac{1}{2} P_z \mu'  = 0. \label{Bianchi2}
\end{equation}
With the $h(r)$ function inserted, (\ref{Bianchi}) becomes
\begin{equation}
 \frac{f'}{f} - h \mu'  = 0. \label{Bianchi3}
\end{equation}

\section{General calculations from the field equations} \label{main}

In Paper 1, it has been shown that, combining (\ref{G00}) with (\ref{G03}), one obtains
\begin{equation}
\left(\frac{kf' - fk'}{D}\right)' = 0, \label{sol1}
\end{equation}
which can be integrated as \cite{D06}
\begin{equation}
kf' - fk' = 2c D, \label{sol2}
\end{equation}
where $2c$ is an integration constant and where the factor 2 is chosen for further convenience.

Then, using (\ref{D2}) into (\ref{sol2}) one obtains
\begin{equation}
\frac{f'l'+k'^2}{2D^2} = \frac{f'D'}{fD} - \frac{f'^2}{2f^2} + \frac{2c^2}{f^2}, \label{sol3}
\end{equation}
which, after being inserted into (\ref{G11}), gives
\begin{equation}
\left(\mu' + \frac{f'}{f} \right) \frac{D'}{D} - \frac{f'^2}{2f^2} + \frac{2c^2}{f^2} =0, \label{sol4}
\end{equation}
that becomes, with (\ref{Bianchi3}) inserted,
\begin{equation}
\frac{D'}{D} = \frac{h}{1+h} \left(\frac{f'}{2f} - \frac{2c^2}{ff'} \right). \label{sol5}
\end{equation}

Now, (\ref{G11}) added to (\ref{G22}) yields
\begin{equation}
\frac{D''}{D} = \kappa \rho h \textrm{e}^{\mu}. \label{sol6}
\end{equation}
Then, (\ref{G11}) and (\ref{sol6}) inserted into (\ref{G00}) give
\begin{equation}
-h \mu'' - 2(1+h) \frac{D''}{D} + h\frac{f''}{f} - \left( h\frac{f'}{f} + 3 h \mu' \right) \frac{D'}{D} = 0.  \label{sol7}
\end{equation} 
The Bianchi identity (\ref{Bianchi3}) differentiated with respect to $r$ becomes
\begin{equation}
h \mu'' = \frac{f''}{f} - \frac{f'^2}{f^2} - \frac{h'f'}{hf}.  \label{sol8}
\end{equation} 
Then, one can insert (\ref{Bianchi3}) and (\ref{sol8}) into (\ref{sol7}) such as to obtain
\begin{eqnarray}
&-& (1-h)\frac{f''}{f} - 2(1+h)\frac{D''}{D} + \frac{f'^2}{f^2} \nonumber \\
&+& \frac{h'f'}{hf} -(3+h)\frac{f'D'}{fD}=0. \label{sol9}
\end{eqnarray}
Using the identity
\begin{equation}
\frac{D''}{D} = \left(\frac{D'}{D}\right)' + \left( \frac{D'}{D}\right)^2,  \label{sol10}
\end{equation} 
$D''/D$ can be calculated from (\ref{sol5}) as
\begin{eqnarray}
\frac{D''}{D} &=& \frac{h'}{(1+h)^2} \left( \frac{f'}{2f} - \frac{2c^2}{ff'} \right) \nonumber\\
&+&\frac{h}{1+h} \left( \frac{f''}{2f} - \frac{f'^2}{2f^2} + \frac{2c^2 f''}{ff'^2} + \frac{2c^2}{f^2} \right)  \nonumber\\
&+& \frac{h^2}{(1+h)^2} \left(\frac{f'}{2f} - \frac{2c^2}{ff'} \right)^2, \label{sol11}
\end{eqnarray}
and inserted into (\ref{sol9}) together with (\ref{sol5}) such as to obtain
\begin{equation}
\left(4hc^2 + f'^2 \right) \left[\frac{4c^2}{f^2} + \frac{2(1+h)}{h} \frac{f''}{f} - \frac{2h'}{h^2} \frac{f'}{f} - \frac{2+h}{h}\frac{f'^2}{f^2} \right] = 0. \label{sol12}
\end{equation}
Two possibilities can be distinguished at this stage. One corresponds to the vanishing of the first factor in (\ref{sol12}), and the other to the vanishing of the second factor.

The vanishing of the first factor can be written as
\begin{equation}
f'^2 = -4hc^2, \label{I1}
\end{equation}
which implies $h<0$. The weak energy condition $\rho>0$ imposes therefore, $P_z < 0$. However, since a negative pressure is acceptable for different applications, this case will be considered throughout.

Dividing  (\ref{I1}) by the Bianchi identity (\ref{Bianchi3}), one obtains
\begin{equation}
ff' = - \frac{4c^2}{\mu'}, \label{I2}
\end{equation}
which can be inserted into (\ref{sol5}) together with (\ref{I1}) and give
\begin{equation}
h \mu' =   \frac{2 D'}{D}. \label{I3}
\end{equation}
Now, using (\ref{Bianchi3}) into (\ref{I3}), it comes
\begin{equation}
\frac{f'}{f} =  \frac{2 D'}{D}, \label{I4}
\end{equation}
which can be integrated as
\begin{equation}
f =  c_f D^2, \label{I5}
\end{equation}
where $c_f$ is an integration constant.

Now, (\ref{I5}) differentiated with respect to $r$ and the result inserted into (\ref{I1}), then divided by (\ref{I5}), yields
\begin{equation}
\frac{D'}{D} =\frac{c\sqrt{-h}}{f}, \label{I6}
\end{equation}
which is then inserted into (\ref{sol5}) to give a second degree in $f'$ equation that reads
\begin{equation}
f'^2 - \frac{2c \sqrt{-h}(1+h)}{h} f' - 4c^2 = 0, \label{I7}
\end{equation}
whose solutions are
\begin{equation}
f' =  \frac{c \sqrt{-h}(1+h)}{h}  + \epsilon c \sqrt{4 - \frac{(1+h)^2}{h}}, \label{I8}
\end{equation}
where $\epsilon = \pm 1$ and which is inserted into (\ref{I1}) to obtain
\begin{equation}
\left[c \sqrt{-h}\frac{(1+h)}{h} + \epsilon c \sqrt{4-\frac{(1+h)^2}{h}} \right]^2 =  -4hc^2, \label{I9}
\end{equation}
that is a polynomial equation in $h$ with constant coefficients, so that the roots, if they exist, are constant values for $h$. Now, it has been shown by C\'el\'erier and Santos \cite {CS20} that a constant value for $h$ implies a trivial Minkowski spacetime. Therefore, this case is ruled out and only the vanishing of the second factor in (\ref{sol12}) will now be considered.

\section{General method for constructing exact solutions to this problem} \label{method}

The key tools of this method are the two auxiliary functions defined above. The function $D$, defined by $D^2=fl+k^2$, was introduced by van Stockum \cite{vS37} for his construction of the metric corresponding to a rigidly rotating dust cylinder matched to the Lewis family of vacuum exteriors. This function has been used since then in a series of works dealing with stationary cylindrical fluids in GR \cite{H98,D06,C19,CS20}. It is deprived of any physical meaning, but rather easy to integrate. It is used here to obtain in particular the $l$ metric function once the others are found without having to bother with integrating the fastidious (\ref{G33}) field equation.

The second auxiliary function is the ratio $h$ of the energy density of the fluid over its pressure. It is the new and key tool of the present series of works. It was implicitly introduced by Debbasch et al. \cite{D06} as a constant parameter $\alpha$. Here, it is allowed to vary with the radial coordinate $r$ and becomes the function $h(r)$. Actually, after integration, the field equations yield the metric and the different physical quantities of interest as functions of $h$. An implicit or explicit expression is found for $h(r)$ itself, depending on the class of solutions. This tool is a strong improvement for the finding of exact solutions to these series of configurations.

Going back to the equations to be solved, their solutions satisfy the vanishing of the second factor in (\ref{sol12}), i. e.,
\begin{equation}
\frac{4c^2}{f^2} + \frac{2(1+h)}{h} \frac{f''}{f} - \frac{2h'}{h^2} \frac{f'}{f} - \frac{2+h}{h}\frac{f'^2}{f^2} = 0. \label{II1}
\end{equation}

However, up to now, only five independent differential equations have been made available for six unknowns, i.e., the four metric functions, $f$, $k$, $\textrm{e}^{\mu}$, $l$, the energy density $\rho$ and the pressure $P_z$, alternatively the ratio $h$. Therefore, the set of equations needs to be closed by an additional constraint which is chosen here, for mathematical purpose, to be the fixing of an expression for $f$ as a function of $h$. At the end of the calculations, this constraint transforms into properties of the physical features of the fluid, which can be selected for further physical applications.

Once $f(h)$ is set, the derivatives of this function with respect to $r$ are replaced in (\ref{II1}) by
\begin{equation}
f' = \frac{\textrm{d}f}{\textrm{d}h}h', \label{II2}
\end{equation}
\begin{equation}
f'' = \frac{\textrm{d}f}{\textrm{d}h}h'' + \frac{\textrm{d}^2f}{\textrm{d}h^2}h'^2, \label{II3}
\end{equation}
that gives
\begin{equation}
h'' + \left[\frac{\frac{\textrm{d}^2f}{\textrm{d}h^2}}{\frac{\textrm{d}f}{\textrm{d}h}} - \frac{(2+h)}{2(1+h) f} \frac{\textrm{d}f}{\textrm{d}h} - \frac{1}{h(1+h)} \right] h'^2 + \frac{2c^2 h}{(1+h) f \frac{\textrm{d}f}{\textrm{d}h}}= 0. \label{II4}
\end{equation}

Denoting
\begin{equation}
A(h) \equiv \frac{\frac{\textrm{d}^2f}{\textrm{d}h^2}}{\frac{\textrm{d}f}{\textrm{d}h}} - \frac{(2+h)}{2(1+h) f} \frac{\textrm{d}f}{\textrm{d}h} - \frac{1}{h(1+h)}, \label{II5}
\end{equation}
\begin{equation}
B(h) = \frac{2c^2 h}{(1+h) f \frac{\textrm{d}f}{\textrm{d}h}}, \label{II6}
\end{equation}
the general solution of (\ref{II4}) reads
\begin{equation}
r = \epsilon \int _{h_0} ^h \frac{\exp\left( \int_1 ^{\theta} A(v) \textrm{d}v \right)}{ \sqrt{-2 \int_1^{\theta} \exp\left(2 \int_1^u A(v) \textrm{d}v \right)B(u) \textrm{d}u}} \textrm{d} \theta, \label{II7}
\end{equation}
which implies
\begin{equation}
h' = \epsilon \frac{\sqrt{-2 \int_0^h \exp\left(2 \int_1^u A(v) \textrm{d}v \right)B(u) \textrm{d}u}}{\exp\left( \int_0 ^h A(v) \textrm{d}v \right)}, \label{II8}
\end{equation}
which yield explicit expressions of $h$ for $r$ and $h'$, provided both integrals can be integrated, that depends on the form of $f(h)$.

Inserting (\ref{II1}) into (\ref{sol5}), one obtains
\begin{equation}
\frac{D'}{D} = \frac{f''}{f'} - \frac{f'}{(1+h)f} - \frac{h'}{h(1+h)}, \label{II9}
\end{equation}
where $f'$ is replaced by its expression (\ref{II2}) in order to obtain
\begin{equation}
\frac{D'}{D} = \frac{f''}{f'} - \frac{h'}{h(1+h)} - \frac{1}{(1+h)f}\frac{\textrm{d}f}{\textrm{d}h}h', \label{II10}
\end{equation}
which can be integrated as
\begin{equation}
D =c_D \frac{(1+h)}{h}f' \exp\left(-\int_{h_0}^h \frac{1}{(1+h)f}\frac{\textrm{d}f}{\textrm{d}h}\textrm{d}h \right), \label{II11}
\end{equation}
with $f'$ given by (\ref{II2}) and thus known once $f(h)$ is chosen and with $h'(h)$ given by (\ref{II8}). An explicit expression for $D(h)$ is obtained if the integral in the exponential can be integrated.

If $f(h)$, $D(h)$ and $h'(h)$ are explicitly known as it might appear at this stage, one can calculate the metric function $k$ as the solution of the differential equation (\ref{sol2}) that reads
\begin{equation}
k = f \left( c_k - 2c \int_0^r \frac{D(v)}{f(v)^2} \textrm{d}v \right), \label{II12}
\end{equation}
where $c_k$ is an integration constant. A change of integration variable $r \rightarrow h$ implies then
\begin{equation}
k = f \left( c_k - 2c \int_{h_0}^h \frac{D(v)}{f(v)^2 h'(v)} \textrm{d}v \right), \label{II13}
\end{equation}
which gives an explicit expression for $k(h)$, provided the integral can be integrated. Now, $l$
follows from the definition of $D^2$ given by (\ref{D2}), i.e.,
\begin{equation}
l = \frac{D^2 - k^2}{f}. \label{II14}
\end{equation}

Then, the last metric function $\textrm{e}^{\mu}$ can be obtained from the Bianchi identity (\ref{Bianchi3}) written as
\begin{equation}
\mu' = \frac{f'}{hf}, \label{II15}
\end{equation}
where one inserts $f'$ given by (\ref{II2}), so that this equation can be integrated as
\begin{equation}
\textrm{e}^{\mu} = \exp \left(\int_{h_0}^h\frac{1}{hf}\frac{\textrm{d}f}{\textrm{d}h} \textrm{d}h \right), \label{II16}
\end{equation}
which yields an explicit expression for $\textrm{e}^{\mu} (h)$, provided the integral can be explicitly integrated.

Finally, the energy density $\rho$ proceeds from the addition of (\ref{G11}) and (\ref{G22}) that gives
\begin{equation}
\frac{D''}{D} = \kappa h\rho \textrm{e}^{\mu}, \label{II17}
\end{equation}
with $D''$ written as
\begin{equation}
D'' = \frac{\textrm{d}\left(\frac{\textrm{d}D}{\textrm{d}h}h'\right)}{\textrm{d}h} h', \label{II18}
\end{equation}
where $h'(h)$ given by (\ref{II8}) is inserted to obtain $D''(h)$ which is substituted into (\ref{II18}) so as to obtain $\rho$ that reads
\begin{equation}
\rho = \frac{1}{\kappa h} \frac{D''(h)}{D(h) \textrm{e}^{\mu}(h)}, \label{II19}
\end{equation}
from which $P_z$ comes easily, considering the definition of $h$ implying $P_z = h \rho$, as
\begin{equation}
P_z = \frac{1}{\kappa} \frac{D''(h)}{D(h) \textrm{e}^{\mu}(h)}. \label{II20}
\end{equation}

Now, the axisymmetry of the cylindrical spacetime imposes a condition on the metric that reads \cite{S09,P96}
\begin{equation}
l \stackrel{0}{=} 0, \label{II21}
\end{equation}
with $\stackrel{0}{=}$ denoting values taken at the rotation axis where $r=0$.
The regularity condition, which is meant to ensure Lorentzian geometry, i. e., 'elementary flatness', i.e., the standard $2\pi$--periodicity of the $\phi$ coordinate, in the vicinity of the axis, can be written as \cite{S09,M93}
\begin{equation}
\frac{l_{,\alpha}l^{,\alpha}}{4l} \stackrel{0}{=} 1, \label{II22}
\end{equation}
which, in the case of metric (\ref{metric}), becomes \cite{D06}
\begin{equation}
\frac{\textrm{e}^{-\mu} l'^2}{4l} \stackrel{0}{=} 1. \label{II23}
\end{equation}
These conditions provide two constraints on the integration constants of the metric, reducing by two the number of independent parameters of the solutions. Note, however, that in some cases where, in particular, the metric does not diverge at the axis, the regularity condition can represent a nuisance for an otherwise well-behaved solution. Indeed, this condition which is sufficient but not necessary, can add needless constraints on the parameters, thus inducing ill-behaved characteristics for the metric. An example is given in Appendix \ref{ApB}.

Moreover, owing to the form of the solutions obtained with different realizations of $f(h)$, some of which are displayed below, it seems generic that a rescaling of the $\phi$ coordinate can be done, such as to reduce once more the number of independent parameters.

Now, to be of use in an astrophysical context, the solutions describing the interior of the fluid must be matched, on the boundary $\Sigma$, to a physically relevant exterior. It has been given in Paper 1 the reasons why the Lewis-Weyl vacuum spacetime is perfectly suitable for this purpose. Using this exterior solution and applying it to the junction condition displayed by Israel \cite{I66}, a constraint on the principal stresses of the fluid has been obtained by Debbasch et al. \cite{D06} and reads $P_r \stackrel{\Sigma}{=} 0$. This is generically satisfied by the present fluids whose equation of state imposes $P_r=0$ on the whole spacetime.

Finally, different physical and mathematical properties of the solutions thus obtained can be calculated and analyzed \cite{C21}.

Any expression for $f(h)$ leading to the possibility of integrating all the above integrals is the seed for an exact solution describing those spacetimes. A particular example of such a solution has been displayed in Paper 1. However, at variance with what has been advocated there, it must be stressed that the choice of a given $f(h)$, or equivalently of $\mu(h)$, is not a gauge choice but corresponds to considering a particular physical system whose characteristics can be described by, e. g., the issuing expressions for $\rho$ or $P_z$.

\section{Reminder, summary and addenda concerning Paper 1 solutions} \label{paper1}

A concise reminder of the solution described in Paper 1 is provided here to allow the reader an easier comparison of the three classes of solutions displayed for the purely axial pressure case. Moreover, additional clarifications are also provided in this section for a better characterisation of some features of these solutions. It is easy to verify that the class  proposed in Paper 1 satisfies indeed (\ref{II1}). Recalling that these solutions imply, inter alias,
\begin{equation}
f = \left( \frac{1-h_0}{1-h}\right)^2, \label{II24}
\end{equation}
\begin{equation}
h'= \left\{\frac{h^2(1-h)^3}{1+h}\left[ 2\ln \frac{h_0}{h} + 4 (h-h_0) - (h^2 - h_0^2)\right]\right\}^\frac{1}{2}, \label{II25}
\end{equation}
and inserting (\ref{II24}) and (\ref{II25}) into (\ref{II1}), one can verify that they satisfy (\ref{II1}), provided
\begin{equation}
c^2 = (1-h_0)^4, \label{II26}
\end{equation}
showing that the solutions depend on only one parameter, $h_0$, the value of the ratio $h$ on the axis. As it has been stressed in Paper 1 and is demonstrated in Appendix A of the present article, the absolute value of the parameter $c$ gives the amplitude of the rotation scalar of the fluid on the axis. Therefore, this amplitude also depends only on $h_0$. Hence, the interpretation of the $c$ parameter of the exterior Lewis spacetime follows, for this configuration of the source, as depending only on the value of the ratio of the pressure over the energy density evaluated on the axis.

Inserting (\ref{II26}) into the metric functions and other expressions describing the physical properties of these solutions as displayed in Paper 1, one obtains their final form as
\begin{equation}
f = \left( \frac{1-h_0}{1-h}\right)^2, \label{II27}
\end{equation}
\begin{equation}
\textrm{e}^{\mu} = \left(\frac{1-h_0}{h_0}\right)^2 \left(\frac{h}{1-h}\right)^2, \label{II28}
\end{equation}
\begin{equation}
k = (1-h_0)^2 \frac{\left[ 2\ln \frac{h_0}{h} + 4 (h-h_0) - (h^2 - h_0^2)\right]}{(1-h)^2}, \label{II29}
\end{equation}
\begin{eqnarray}
&&l =  (1-h_0)^2\left[ 2\ln \frac{h_0}{h} + 4 (h-h_0) - (h^2 - h_0^2)\right] \nonumber \\
&\times& \left\{\frac{1-h}{1+h} - \frac{\left[ 2\ln \frac{h_0}{h} + 4 (h-h_0) - (h^2 - h_0^2)\right]}{(1-h)^2}\right\}, \label{II30}
\end{eqnarray}
\begin{equation}
r = \epsilon\int_{h_0}^{h} \left\{\frac{1+u}{u^2(1-u)^3\left[ 2\ln \frac{h_0}{u} + 4 (u-h_0) - (u^2 - h_0^2)\right]}\right\}^\frac{1}{2} \textrm{d}u, \label{II31}
\end{equation}
where $\epsilon = \pm 1$ exhibits the same sign as $c$, due to the extraction of both square roots in (\ref{II26}).
\begin{equation}
h' = \epsilon \left\{\frac{h^2(1-h)^3}{1+h}\left[ 2\ln \frac{h_0}{h} + 4 (h-h_0) - (h^2 - h_0^2)\right]\right\}^{\frac{1}{2}}, \label{II32}
\end{equation}
\begin{eqnarray}
&&\rho = \frac{2 h_0^2}{\kappa} \frac{(1-h)^4}{h^2 (1+h)^2} \nonumber \\
&\times& \left\{\frac{h\left[ 2\ln \frac{h_0}{h} + 4 (h-h_0) - (h^2 - h_0^2)\right]}{(1+h)} +  2(1-h)^2 \right\}. \label{II33}
\end{eqnarray}
\begin{equation}
\omega^2 = \frac{h_0^2}{(1 - h_0)^2} \frac{(1 - h)^6}{h^2}, \label{II34}
\end{equation}
\begin{equation}
\omega^2 \stackrel{0}{=} (1 - h_0)^4.
\label{II35}
\end{equation}

Two cases have been considered in Paper 1. 

Case (i) fulfils both weak and strong energy conditions, i.e., $h>0$. For the subcase $\epsilon =-1$, $h(r)$ is monotonically decreasing from $h_0$ on the axis down to $h_{\Sigma}$ at the boundary, which verify

$0<h_1<h_{\Sigma}<h<h_0<h_2<+1$, where $h_1$ and $h_2$ are determined by (95) and (94) of Paper 1.

For the subcase $\epsilon = +1$, $h(r)$ is monotonically increasing according to the following inequalities:

$0<h_1<h_0<h<h_{\Sigma}<h_2<+1$.

The second case, denoted (ii), exhibits a behaviour independent of the sign of $\epsilon$. It does not satisfy the strong energy condition since $h<0$ here, but it does verify the weak energy one. The $h(r)$ function evolves according to

$-1<h_0<h<h_{\Sigma}<h_3<0$, with $h_3$ defined by (96) of Paper 1.

Now, one could wish a more detailed analysis of the behaviour of the functions expressing the physical properties of this fluid. However, such features are extremely dependent on the values of $h_0$. It is actually easy to verify that, even for solutions belonging to the same case and subcase, the behaviour of, e. g., the energy density and the pressure, can differ substantially depending on the situation of $h_0$ inside its allowed range. This is not the case for, e. g., the class A solutions displayed in below Section \ref{A}, whose energy density and pressure behaviours are determined independently of $h_0$, save for a constant factor. However, since the purpose of this work is to provide a method for finding a metric suitable for representing (astro)physical objects or regions, the fact that such behaviours are utterly dependent on the value of the parameter $h_0$ should be considered as a benefit. Indeed, once a class of solutions has been selected owing to some desired particular features, this allows a tuning of the actual solution to the wanted physical properties through different trials of the $h_0$ value, which is easier when the choice is wider.

In Paper 1, three possible singularities have been exhibited and discussed for this class of solutions. The loci were $h=+1$, $h=-1$ and $h = 0$. To enhance the results of the analysis displayed there, the Kretschmann scalar $K$ have been computed using the SageMath \cite{S21} and Mathematica \cite{M22} softwares. A simplified expression is reproduced below, with a focus on the main parts of interest. It is therefore written as
\begin{equation}
K = \frac{4(1-h)^6 h_0^4 Y(h_0,h)}{h^4(1+h)^6(1-h_0)^4 \left[2\ln\frac{h_0}{h}+4(h-h_0)-(h^2-h_0^2)\right]^4},
\label{K1}
\end{equation}
where $Y(h_0,h)$ denotes a function of $h$ implying $h_0$ as a parameter that is too long to be displayed here while it is anyhow useless for the present purpose. It is easy to see that $K$ vanishes for $h = +1$. From the expression of the denominator, it comes that $K$ diverges for $h = -1$, $h=0$ and $h=h_0$. This solves the issue of the characterization of the singularity at $h=+1$ which can definitely be considered as a mere coordinate singularity as suggested in Paper 1. At two other singular loci, where $h= -1$, and $h=0$, the energy density of the fluid happens to diverge, as stressed in Paper 1. Therefore, coupled to the diverging of the curvature invariant $K$, this implies that for these values of the ratio $h$ one is confronted to curvature singularities. However, since the energy density diverges at locations which constitute limiting values for the ratio pressure over energy density, the spacetime cannot be extended beyond those points and therefore the singularity is not too awkward. 

Now, the diverging of $K$ at the axis is more unexpected. Indeed, in Paper 1, the solution has been shown to fulfil regularity and ''elementary flatness'' conditions, while the behaviour of $K$ points to a curvature singularity at the axis. This might be an additional hint in favor of the weakness of regularity conditions \cite{L94,W96,C00}.

\section{Exemplifying the general method with a new fully integrated solution} \label{example}

Now, the solution displayed in Paper 1, even though sufficiently characterized to allow a thorough analysis of its mathematical and physical properties, lacks an explicit expression for $h(r)$ which is merely depicted by its first derivative with respect to $r$, i.e., $h'(h)$. Therefore, a more complete solution is displayed and analyzed in this section. The purpose is twofold: first, to exemplify the use of the method described in Section \ref{method} which is different and more general than the one employed in Paper 1, second, to describe more thoroughly a fully integrated solution which will allow one to exhibit the permanent features of this physical configuration. As it will be shown, this solution splits into two classes, depending on their ability to verify one or the other conditions usually ascribed to such systems. The properties common to both classes will therefore be displayed first.

\subsection{Common features} \label{common}

First, a simple expression is assigned to the $f(h)$ function that reads
\begin{equation}
f = \frac{h_0}{h}, \label{II27}
\end{equation}
which yields
\begin{equation}
\frac{f'}{f} = - \frac{h'}{h}, \label{II28}
\end{equation}
\begin{equation}
\frac{f''}{f} = - \frac{h''}{h} +\frac{2h'^2}{h^2}. \label{II29}
\end{equation}
The three equations (\ref{II27})-(\ref{II29}) are inserted into (\ref{II1}) which gives
\begin{equation}
h'' - \frac{(4+3h)}{2h(1+h)}h'^2 - \frac{2c^2}{h_0^2}\frac{h^4}{(1+h)} = 0. \label{II30}
\end{equation}
The solution (\ref{II7}) of (\ref{II30}) can then be integrated as
\begin{equation}
r = \epsilon \frac{2}{3} \left[\left(\frac{1+h}{h}\right)^{\frac{3}{2}} - \left(\frac{1+h_0}{h_0}\right)^{\frac{3}{2}} \right], \label{II31}
\end{equation}
from which 
\begin{equation}
h' = -\epsilon \frac{2c}{h_0} \frac{h^{\frac{5}{2}}}{(1+h)^{\frac{1}{2}}} \label{II32}
\end{equation}
follows.

Now, (\ref{II27})-(\ref{II29}) are inserted into (\ref{II9}) to obtain
\begin{equation}
\frac{D'}{D} = \frac{h''}{h'} -\frac{2h'}{h}, \label{II33}
\end{equation}
which can be integrated as
\begin{equation}
D = c_D\frac{h'}{h^2}, \label{II34}
\end{equation}
where $c_D$ is an integration constant. Then, $h'$ given by (\ref{II32}) is inserted into (\ref{II34}) and yields
\begin{equation}
D = - \epsilon \frac{2c c_D}{h_0}\sqrt{\frac{h}{1+h}}. \label{II35}
\end{equation}

Now, to calculate the metric function $k$, using (\ref{II13}), the following expression is needed
\begin{equation}
\frac{D}{f^2 h'} = \frac{c_D}{h_0^2}, \label{II36}
\end{equation}
that gives, after being inserted into (\ref{II13}),
\begin{equation}
k = f \left( c_k - 2c \int_{h_0}^h \frac{c_D}{h_0^2} \textrm{d}v \right). \label{II37}
\end{equation}
Substituting (\ref{II27}) in the above equation and integrating, one obtains
\begin{equation}
k = - \frac{2c c_D}{h_0} + \frac{2 c c_D + h_0 c_k}{h}. \label{II38}
\end{equation}

Now, the metric function $l$ proceeds from (\ref{II14}) where (\ref{II27}), (\ref{II35}) and (\ref{II38}) are inserted and that yields
\begin{equation}
l = - \frac{4c^2 c_D^2}{h_0^3}\frac{h}{1+h} - \frac{(2 c c_D + h_0 c_k)^2}{h_0h} + \frac{4c c_D}{h_0^2}(2c c_D + h_0 c_k). \label{II39}
\end{equation}

Then, from the Bianchi identity (\ref{Bianchi3}), the first derivative of the metric function $\mu$ can be written as
\begin{equation}
\mu' = \frac{f'}{hf}, \label{II40}
\end{equation}
which, after substituting (\ref{II28}), then integrating, becomes
\begin{equation}
\textrm{e}^{\mu} = c_{\mu}\textrm{e}^{\frac{1}{h}}, \label{II41}
\end{equation}
where $c_{\mu}$ is an integration constant.

The energy density $\rho$ proceeds from (\ref{II19}) where (\ref{II35}) and (\ref{II41}) are inserted to yield
\begin{equation}
\rho = \frac{4c^2}{\kappa c_{\mu} h_0^2} \frac{h^2}{(1+h)^3 \textrm{e}^{\frac{1}{h}}}, \label{II42}
\end{equation}
from which $P_z$ follows as
\begin{equation}
P_z = \frac{4c^2}{\kappa c_{\mu} h_0^2} \frac{h^3}{(1+h)^3 \textrm{e}^{\frac{1}{h}}}. \label{II43}
\end{equation}

Now, (\ref{II31}) can be written as
\begin{equation}
h = \frac{1}{\left[\left(\frac{1+h_0}{h_0}\right)^{\frac{3}{2}} + \frac{3 \epsilon r}{2}\right]^{\frac{2}{3}} -1}. \label{II44}
\end{equation}
Differentiating (\ref{II44}) with respect to $r$ and setting 
$r=0$ one obtains the value $h'_0$ of $h'$ on the axis that reads
\begin{equation}
h'_0 = - \frac{\epsilon h_0^{\frac{5}{2}}}{\sqrt{1+h_0}}. \label{II45}
\end{equation}
Another expression for $h'_0$ can be obtained by setting $h=h_0$ into (\ref{II32}), that yields
\begin{equation}
h'_0 = - 2\epsilon c \frac{ h_0^{\frac{3}{2}}}{(1+h_0)^{\frac{1}{2}}}. \label{II46}
\end{equation}
Comparing (\ref{II45}) and (\ref{II46}), one obtains
\begin{equation}
2 c = h_0. \label{II47}
\end{equation}

Now, the axisymmetry and regularity conditions have to be implemented. The axisymmetry condition, (\ref{II21}), implies
\begin{equation}
\frac{4c^2 c_D^2}{c_k^2} = h_0(1+h_0), \label{II48}
\end{equation}
while the regularity condition (\ref{II22}), with (\ref{II48}) inserted, yields
\begin{equation}
h_0 = - \frac{4}{3}, \label{II49}
\end{equation}
which is consistent with the axisymmetry condition since, the expression (\ref{II49}) of $h_0$ substituted into (\ref{II48}) gives
\begin{equation}
c_k = 3 c c_D. \label{II50}
\end{equation}

However, when the weak energy condition, $\rho>0$, is fulfilled, (\ref{II49}) is inconsistent with the strong energy condition which would then impose $P_z>0$. Therefore, the set of solutions can be splitted in two classes: class A, including the spacetimes that satisfy the weak and strong energy conditions, but not the regularity condition, and class B, composed of a single solution satisfying the weak energy and the regularity conditions, but not the strong energy one.

Now, the regularity condition, not satisfied by class A, is obtained by a normalisation of the $\phi$ coordinate provided elementary flatness is assumed in the vicinity of the axis. However, elementary flatness by no means guarantees regularity \cite{L94}. Moreover, regularity being defined in Cartesian type coordinates, a violation of the regularity condition might be linked to some issue with the use of polar type coordinates \cite{W96}. The adaptation of polar type coordinates to the study of features of the axis has also been discussed by Carot \cite{C00}, while the presence of some matter on the axis has been proposed to deal with the nonregularity problem \cite{P96}. Indeed, a singularity on the axis seems not to be a real problem provided it should be no worse than conical. A conical singularity occurs on the axis if the limiting value of the expression $\frac{\textrm{e}^{-\mu} l'^2}{4l}$ is finite and nonvanishing there. It is indeed the case here, where the axial singularity is therefore merely conical.

The nonfulfilment of the strong energy condition, implying $P_z<0$, does not either exclude as such the solution of class B. Indeed, such a negative pressure, which is inconsistent when standard states of fluids are considered, can appear in other physical and cosmological circumstances which will be discussed in the conclusion.

\subsection{Class A} \label{A}

Two constraints have to be implemented into the solutions of this subclass: (\ref{II48}) coming from axisymmetry which allows $c_k$ to be replaced by $2 c c_D/ \sqrt{h_0(1+h_0)}$, the sign indeterminacy being absorbed into the definitions of the constants and (\ref{II47}) which is used to replace $c$ by $h_0/2$. After completion of the first replacement, the metric functions $k$ and $l$ become
\begin{equation}
k = 2 c  c_D \left[\left(1+\sqrt{\frac{h_0}{1+h_0}}\right) \frac{1}{h} - \frac{1}{h_0}\right], \label{II51}
\end{equation}
\begin{eqnarray}
l &=& 4 c^2 c_D^2\left[\frac{2}{h_0^2}\left(1+\sqrt{\frac{h_0}{1+h_0}}\right) - \frac{h}{h_0^3(1+h)} \right. \nonumber \\
&-& \left. \left(1+\sqrt{\frac{h_0}{1+h_0}} \right)^2 \frac{1}{h_0h} \right]. \label{II52}
\end{eqnarray}
Noticing that $2cc_D$ appears as such in the expression (\ref{II51}) for $k$ and squared in the expression (\ref{II52}) for $l$, one can make a rescaling of the $\phi$ coordinate from a factor $2cc_D$, which implies setting $2cc_D=1$ so that the upper limit of definition of $\phi$ remains $2\pi$.

Then, after completing the three above described replacements into every expression of interest, the solution can be summarized as
\begin{equation}
f = \frac{h_0}{h}, \label{II53}
\end{equation}
\begin{equation}
\textrm{e}^{\mu} = c_{\mu}\textrm{e}^{\frac{1}{h}}, \label{II54}
\end{equation}
\begin{equation}
k = \left(1+\sqrt{\frac{h_0}{1+h_0}}\right) \frac{1}{h} - \frac{1}{h_0}, \label{II55}
\end{equation}
\begin{equation}
l = \frac{2}{h_0^2}\left(1+\sqrt{\frac{h_0}{1+h_0}}\right) - \frac{h}{h_0^3(1+h)} - \left(1+\sqrt{\frac{h_0}{1+h_0}} \right)^2 \frac{1}{h_0h}, \label{II56}
\end{equation}
\begin{equation}
\rho = \frac{h_0^2}{\kappa c_{\mu}} \frac{h^2}{(1+h)^3 \textrm{e}^{\frac{1}{h}}}, \label{II57}
\end{equation}
\begin{equation}
P_z = \frac{h_0^2}{\kappa c_{\mu}} \frac{h^3}{(1+h)^3 \textrm{e}^{\frac{1}{h}}}, \label{II58}
\end{equation}
\begin{equation}
D = - \frac{\epsilon}{h_0}\sqrt{\frac{h}{1+h}}, \label{II59}
\end{equation}
\begin{equation}
h = \frac{1}{\left[\left(\frac{1+h_0}{h_0}\right)^{\frac{3}{2}} + \frac{3 \epsilon r}{2}\right]^{\frac{2}{3}} -1}. \label{II59}
\end{equation}

Now, a set of significant mathematical and physical properties of this class are going to be analyzed. Their behaviour depends strongly on the sign of $\epsilon$.

\subsubsection{Sign constraints and metric signature} \label{sign}

Class A solutions verify, by definition, the strong energy condition which implies $\rho>0$ and $P_z \geq 0$, hence $h\geq0$. It is easy to verify from (\ref{II57})  that $\rho>0$ implies $c_{\mu}>0$ , which reinforce the above argument for a positive $c_{\mu}$. The positiveness of $P_z$ given by (\ref{II58}) follows from an analogous argument. The consistency of these different signs is therefore established.

Now, to obtain a correct Lorentzian signature, $(-+++)$ or $(+---)$, the four metric functions, $f$, $\textrm{e}^{\mu}$, $k$, and $l$, must be all positive or all negative definite. They will therefore be examined in turn from this perspective.

The metric function $f$ as given by (\ref{II53}) is indeed positive definite since $h_0$ and $h$ are positive.

The function $\textrm{e}^{\mu}$, since $c_{\mu} > 0$,
is also positive definite owing to such property of the exponential function.

The metric function $k$, as given by (\ref{II55}), must therefore be positive which implies, using $h>h_0$ as established below in Sec. \ref{bh},
\begin{equation}
h_0 < h < h_0\left(1 + \sqrt{\frac{h_0}{1 + h_0}}\right), \label{II71a}
\end{equation}
which shows that, for the signature to be proper, the $h$ ratio must be bounded by $h_0\left(1 + \sqrt{\frac{h_0}{1 + h_0}}\right)$. This implies that the value $h_{\Sigma}$ of $h$ on the $\Sigma$ boundary cannot be larger than this limit. The ratio $h$ of Class A solutions is therefore increasing from $h_0$ to $h_{\Sigma} < h_0\left(1 + \sqrt{\frac{h_0}{1 + h_0}}\right)$.

The derivative $l'$ of $l$ is obtained by differentiating (\ref{II56}) with respect to $r$, then inserting $h'$ as given by (\ref{II32}), which yields 
\begin{equation}
l' = \frac{-\epsilon}{h_0^3}\sqrt{\frac{h}{1+h}}\left[h_0^2\left(1+\sqrt{\frac{h_0}{1+h_0}}\right)^2 - \frac{h^2}{(1+h)^2}\right]. \label{II72a}
\end{equation}
Implementing (\ref{II71a}) into (\ref{II72a}), one obtains that the sign of $l'$ is the same as that of $-\epsilon$. To obtain $l>0$ in the vicinity of the axis where $l=0$ owing to the axisymmetry condition, $l$ must be increasing with $r$ there, hence $l'(h_0)$ must be positive which implies 
\begin{equation}
\epsilon = -1.
\label{II73a}
\end{equation}
Now, (\ref{II72a}) with (\ref{II73a}) inserted becomes
\begin{equation}
l' = \frac{1}{h_0^3}\sqrt{\frac{h}{1+h}}\left[h_0^2\left(1+\sqrt{\frac{h_0}{1+h_0}}\right)^2 - \frac{h^2}{(1+h)^2}\right], \label{II74a}
\end{equation}
which, for the interval of definition of $h$, (\ref{II71a}), and $h>0$ is positive everywhere. The metric function, increasing from $l(h_0)=0$ to $l(h_{\Sigma})>0$ is therefore positive definite. Hence the signature of the metric is consistent provided $\epsilon$ satisfies (\ref{II73a}).

\subsubsection{Behaviour of the $h(r)$ function} \label{bh}

Since $\epsilon=-1$, the first derivative of $h$ evaluated on the axis, $h'_0$, as given by (\ref{II45}), is positive or null. Because the radial coordinate $r$ is also positive or null, (\ref{II31}) with $\epsilon = -1$ implies
\begin{equation}
\frac{1+h_0}{h_0} \geq \frac{1+h}{h}, \label{II60}
\end{equation}
that gives, owing to the strong energy condition which imposes $h \geq 0$,
\begin{equation}
h \geq h_0, \quad \forall r \geq 0,\label{II61}
\end{equation}
which implies $h$ non-decreasing in the vicinity of the axis $r=0$, that is consistent with $h'_0 \geq 0$. Therefore, with (\ref{II45}) and (\ref{II47}) inserted into (\ref{II32}), it is easy to conclude, since $h_0 > 0$, that $h'$ is positive everywhere and that the function $h$ is monotonically increasing from $h_0$ on the axis to $h_{\Sigma}$ on the boundary, which is consistent with (\ref{II71a}).

\subsubsection{Hydrodynamical properties} \label{hydroI}

The invariant decomposition of the timelike 4-vector $V_{\alpha}$ has been considered in Paper 1 where the different expressions for the acceleration vector, rotation and shear tensor components have been displayed. The nonzero  component of the acceleration vector thus reads
\begin{equation}
\dot{V}_1 = \frac{1}{2} \frac{f'}{f}, \label{II62}
\end{equation}
which becomes, with (\ref{II28}), (\ref{II32}) and (\ref{II47}) inserted,
\begin{equation}
\dot{V}_1 = - \frac{h^{\frac{3}{2}}}{2\sqrt{1+h}}. \label{II63}
\end{equation}
The modulus of this acceleration vector, as displayed in Paper 1, proceeds from
\begin{equation}
\dot{V}^{\alpha}\dot{V}_{\alpha}= \frac{1}{4}\frac{f'^2}{f^2} \textrm{e}^{-\mu}, \label{II65}
\end{equation}
which becomes, with (\ref{II28}), (\ref{II32}), (\ref{II47}) and (\ref{II54}) inserted,
\begin{equation}
\dot{V}^{\alpha}\dot{V}_{\alpha}= \frac{1}{4c_{\mu}}\frac{h^3}{(1+h)} \textrm{e}^{-\frac{1}{h}}. \label{II66}
\end{equation}

Finally, the rotation scalar, whose square has been established in Paper 1 as
\begin{equation}
\omega^2= \frac{c^2}{f^2 \textrm{e}^{\mu}}, \label{II67}
\end{equation}
becomes, with (\ref{II53}) and (\ref{II54}) inserted,
\begin{equation}
\omega^2= \frac{h^2 \textrm{e}^{-\frac{1}{h}}}{4c_{\mu}}. \label{II68}
\end{equation}
At the limit on the axis where $h=h_0$, this rotation scalar takes the value
\begin{equation}
\omega^2 \stackrel{0}{=} \frac{c^2}{c_{\mu}  \textrm{e}^{\frac{1}{h_0}}}, \label{II69}
\end{equation}
where $c$ has been inserted back such as to allow the use of the result displayed in Appendix A where the value of $\omega^2$ at the axis is shown to be equal to $c^2$. Therefore, $c_{\mu} = \textrm{e}^{-{\frac{1}{h_0}}}$. The validity of this result is reinforced by the fact that its expression as an exponential implies a positive $c_{\mu}$. Indeed from the weak energy condition $\rho>0$ and the expression (\ref{II57}) for $\rho$, the sign of $c_{\mu}$ has to be positive, which is consistent with its above determination. Another feature worth to be stressed is that $c$, given here by (\ref{II47}) and by (\ref{II26}) of Paper 1 for the solutions displayed there, is not an independent parameter, as it was put forward too hastily in Paper 1, since it depends actually on $h_0$.

Finally, as it is well-known, rigid rotation implies vanishing shear which applies here of course.

\subsubsection{Singularities} \label{sing}

Notice first that the four metric functions are diverging for $h=0$. However, since $c \neq 0$, such as to be able to account for nonzero values of $\rho$ and $P_z$, and since $c>0$, $h$ satisfying $h \geq h_0 = 2c$ never vanishes. The value $h=0$ is therefore never reached inside these spacetimes.

Then, potentially, the $k$ metric function could vanish for $h = h_0(1+ \sqrt{h_0/(1+h_0)})$. However, it has been shown in Sec. \ref{sign} that $h_{\Sigma} < h_0 (1+ \sqrt{h_0/(1+h_0)})$. Hence this value is never reach by $h$ inside the cylinder.

Finally, the only possible singularity should be located on the axis where $l=0$ and $h=h_0$. However, it has been proven in Sec. \ref{common} that this axial singularity is merely conical and therefore not a real drawback. Moreover, the curvature invariant computed using the SageMath \cite{S21} and Mathematica \cite{M22} softwares, i.e., the Kretschmann scalar, $K$, reads
\begin{eqnarray}
&& K = \frac{\textrm{e}^{\frac{2}{h_0}- \frac{2}{h}}}{4} \left[10h^{\frac{5}{2}}(1+h) + h^3 \sqrt{1+h}\left(-4-37h-68h^2 \right. \right.\nonumber \\
&-& \left. \left. 20h^3+18\sqrt{h(1+h)}+8h^{\frac{3}{2}}\sqrt{1+h}\right)\right](1+h)^{-\frac{13}{2}}, \label{K2}
\end{eqnarray}
which diverges only for $h=-1$ and for $h=h_0$ provided $h_0=-1$. For any other value of $h_0$, the axis is nonsingular. Since a negative value cannot be reached by $h$ in these spacetimes where only strictly positive ratios are allowed, these spacetimes are free of curvature singularities. In particular, the singularity on the axis is confirmed to be soft.

\subsubsection{Behaviour of the energy density and of the pressure} \label{rho}

The derivative with respect to $r$ of the energy density $\rho$ given by (\ref{II57}) reads
\begin{equation}
\rho'= \frac{h_0^2 \textrm{e}^{\frac{1}{h_0}}}{\kappa}\frac{\left[h(2-h)\textrm{e}^{\frac{1}{h}} +1 + h \right]}{(1+h)^4 \textrm{e}^{\frac{2}{h}}} h'. \label{II70}
\end{equation}
Since $h'>0$, the sign of $\rho'$ is that of $h(2-h)\textrm{e}^{\frac{1}{h}} +1 + h$ which has two roots, one negative that is ruled out by the constraint $h>0$, the other positive and approximated by $2.95413$. Therefore, if $h_{\Sigma} < 2.95413$, $\rho$ is monotonically increasing all along from the axis to the boundary $\Sigma$ and, if $h_{\Sigma}>2.95413$, $\rho$ is increasing from the axis where $r=0$ to $r_1$ corresponding to $h_1=2.95413$, then decreasing from $r_1$ down to the boundary.

Now, the derivative with respect to $r$ of the pressure $P_z$ given by (\ref{II58}) reads
\begin{equation}
P'_z= \frac{h_0^2 \textrm{e}^{\frac{1}{h_0}}}{\kappa}\frac{h}{(1+h)^4 \textrm{e}^{\frac{2}{h}}} \left(3h \textrm{e}^{\frac{1}{h}} + 1 + h \right) h'. \label{II71}
\end{equation}
Since $h>0$ and $h'>0$ everywhere, this derivative is also positive and $P_z$ is increasing from the axis to the boundary.

\subsubsection{Final summary of the class A solutions} \label{finalA}

For the reader's convenience a summary of the class A solutions accounting for all the constraints identified in the present section is displayed below
\begin{equation}
f = \frac{h_0}{h}, \label{II72}
\end{equation}
\begin{equation}
\textrm{e}^{\mu} = \textrm{e}^{\frac{1}{h}- \frac{1}{h_0}}, \label{II73}
\end{equation}
\begin{equation}
k = \left(1+\sqrt{\frac{h_0}{1+h_0}}\right) \frac{1}{h} - \frac{1}{h_0}, \label{II74}
\end{equation}
\begin{equation}
l = \frac{2}{h_0^2}\left(1+\sqrt{\frac{h_0}{1+h_0}}\right) - \frac{h}{h_0^3(1+h)} - \left(1+\sqrt{\frac{h_0}{1+h_0}} \right)^2 \frac{1}{h_0h}, \label{II75}
\end{equation}
\begin{equation}
\rho = \frac{h_0^2  \textrm{e}^{\frac{1}{h_0}}}{\kappa} \frac{h^2}{(1+h)^3 \textrm{e}^{\frac{1}{h}}}, \label{II76}
\end{equation}
\begin{equation}
P_z = \frac{h_0^2\textrm{e}^{\frac{1}{h_0}}}{\kappa} \frac{h^3}{(1+h)^3 \textrm{e}^{\frac{1}{h}}}, \label{II77}
\end{equation}
\begin{equation}
D = \frac{1}{h_0}\sqrt{\frac{h}{1+h}}, \label{II78}
\end{equation}
\begin{equation}
h = \frac{1}{\left[\left(\frac{1+h_0}{h_0}\right)^{\frac{3}{2}} - \frac{3 r}{2}\right]^{\frac{2}{3}} -1}, \label{II79}
\end{equation}
\begin{equation}
h'=\frac{h^{\frac{5}{2}}}{\sqrt{1+h}}, \label{II80}
\end{equation}
\begin{equation}
r = \frac{2}{3}\left[\left(\frac{1+h_0}{h_0}\right)^{\frac{3}{2}} - \left(\frac{1+h}{h}\right)^{\frac{3}{2}} \right]. \label{II81}
\end{equation}

\subsection{Class B} \label{B}

This class is defined as including solutions which do not verify the strong but only the weak energy condition, i. e., $\rho>0$ and $P_z<0$, that implies $h<0$. Hence, the regularity condition $h_0=-4/3$ can be satisfied. Inserting this constraint into the axisymmetry condition gives (\ref{II50}), whose substitution into (\ref{II27}), (\ref{II38}), (\ref{II39}) and (\ref{II41}) yields, after implementation of a rescaling of the $\phi$ coordinate from a factor $3cc_D$,
\begin{equation}
f = - \frac{4}{3h}, \label{II82}
\end{equation}
\begin{equation}
\textrm{e}^{\mu} = \textrm{e}^{\frac{1}{h} + \frac{3}{4}}, \label{II83}
\end{equation}
\begin{equation}
k = \frac{1}{2} +  \frac{2}{h}, \label{II84}
\end{equation}
\begin{equation}
l = 3 \left[\frac{1}{2} + \frac{1}{h} + \frac{h}{16(1+h)} \right]. \label{II85}
\end{equation}

Now, the signature of the metric is examined. The $f$ metric function given by (\ref{II82}), is positive definite for negative values of $h$ which are imposed by definition to the solutions of this class. Function $\textrm{e}^{\mu}$ is also everywhere positive owing to a defining property of the exponential function.

However, as shown by (\ref{II84}), $k$ is positive provided
\begin{equation}
h < -4, \label{II91}
\end{equation}
which is not the case for $h_0= -4/3$. Thus, by continuity, $k$ is not positive in a significant region around the axis. Hence, in this region, the signs of $f$, $\textrm{e}^{\mu}$ and $k$ are not consistent. The metric signature is not properly established there. Class B, with the regularity condition imposed, is therefore ruled out.

\section{Paper 1 versus Class A solutions} \label{versus}

Both classes have in common that their metrics and their physical features depend on a single parameter, $h_0$, representing the value of the ratio of the pressure over the energy density measured at the rotation axis. This property is to be contrasted with the presence of four independent parameters in the Lewis solution for the corresponding exterior vacuum. The addition of matter with energy density and pressure reduces the number of degrees of freedom inferred by the three extra parameters in Lewis. A further study of this discrepancy, postponed to future work, might provide new insights into the understanding of stationary rotating spacetimes with cylindrical symmetry and, more generally, into key disparities between vacuum and nonvacuum spacetimes.

Other important features are instead different for both classes. In the Paper 1 solutions, $r(h)$ is given by an implicit integral equation, while, in Class A, $r(h)$ is determined by an explicit expression including the $h_0$ parameter. This allows one to write metric functions and physical properties as explicit expressions of the $r$ coordinate which eases their interpretation but has, as a corollary, the disadvantage of reducing the possibility of adjusting them at will.

Actually, in Paper 1, the behaviours of the functions expressing the physical properties of the gravitating fluid are all depending on the parameter $h_0$ which allows an interesting flexibility in the possibility to reproduce the characteristics of given physical systems. In Class A instead, the hydrodynamical quantities as well as the energy density and the pressure depend on $h_0$ essentially through the function $h(r)$ given by (\ref{II79}) and through a constant overall multiplication factor. They possess therefore less freedom to fit to the properties of any physical structure. In particular, the positive pressure $P_z$ is monotonically increasing from the axis to the boundary whatever the values of $h_0$ and of $h_{\Sigma}$.

Now, in Paper 1, the solutions are divided into three subclasses, one -- (i) with $\epsilon >0$ -- fulfilling the strong energy condition with $h(r)$ increasing from the axis to the boundary, one -- (i) with $\epsilon <0$ -- also fulfilling the strong energy condition but with $h(r)$ decreasing from the axis to the boundary and the last one -- (ii) with no constraint on the sign of $\epsilon$ -- fulfilling only the weak but not the strong energy condition with a negative $h(r)$ increasing from the axis to the boundary. Class A, by contrast, exhibits no subclass. Its solutions satisfy the strong energy but not the regularity condition. However, its axial singularity being conical, this last feature is not a real drawback. Moreover, the allowed range for the ratio $h$ is $0<h_0<h<h_{\Sigma}< h_0(1+\sqrt{h_0/(1+h_0)}$, whose limits are determined by the parameter $h_0$. On the other hand, the solutions displayed in Paper 1 seem to be properly behaving on the axis as they satisfy axisymmetry, regularity and ''asymptotic flatness'' conditions, but they are likely to exhibit a curvature singularity there and also where $h=-1$ and $h=0$, at variance with Class A solutions which are free of such inconveniences. However, since both values are limits of the allowed intervals for $h$, actual Paper 1 spacetimes might be somehow freed from these singularities.

Finally, in both classes, the $h_0$ parameter and the value of the rotation scalar on the axis, $\omega_0$, are strictly related. It reads $\omega_0^2=(1-h_0)^4$ for Paper 1 and $\omega_0^2=h_0^2/4$ for Class A. Therefore, any of both quantities, $\omega_0$ or $h_0$, can be chosen as the parameter defining the solutions in the class, which can be interesting from a physical application point of view.

Recall also that, for a rigidly rotating fluid, the shear vanishes. Hence, this property is shared by both classes of solutions.

\section{Conclusion} \label{concl}

This article is the second of a series devoted to the study of exact solutions for interior spacetimes sourced by a stationary cylinder of fluid rigidly rotating around its symmetry axis and exhibiting an anisotropic pressure. These works constitute the first successful attempt to exhibit exact interior solutions for fluids with anisotropic pressure as gravitational sources in such configurations. For calculation purpose, the configurations have been specialized to three different cases where the pressure is on turn directed alongside each principal stress. The two first articles in the series are devoted to the analysis of the axial pressure case. Indeed, the first axial class found and published in Paper 1 was understood later on to be merely a special case exhibited through a rather complicated calculation. It has been recalled here and its properties revised and supplemented. Then, thanks to the degree of freedom left in the equations, a fully general and simpler method aiming at constructing different classes of solutions of the kind was found. This method has been described in the present paper, named Paper 2, and represents the main result of this work. It is established on the definition and on the use of two auxiliary functions of the $r$ coordinate, $D^2=fl+k^2$, already known for long to be of interest as regards such issue, and a new function $h(r)$ that is the ratio of the pressure nonzero component over the energy density of the fluid. The implementation of such a new tool to carry out the calculations is a key proposal for the finding of the results displayed in this series of articles.

Now, to exemplify the detailed recipe provided here, it has been applied to select two new well behaved classes of solutions, Class A analyzed in Sec. \ref{A}, and another example displayed in Appendix \ref{ApB}. Class A has been shown to verify every condition needing to be satisfied by a fully achieved set of exact solutions, save the regularity condition. The solution of Appendix \ref{ApB} has been provided as an example of a proper metric which can apparently be ruled out by an inappropriate use of the ''regularity condition''.  Since, despite our numerous trials, we were not able to find well behaved solutions verifying both strong energy and regularity conditions, we are led to conjecture that, for this physical configuration, the imposition of the regularity condition might corrupt any well behaved class of solutions. Therefore, the possible issue of an axial singularity has to be solved otherwise, e. g., as shown in Class A.

Besides the calculations used for finding these new solutions, a rather comprehensive analysis of their mathematical and physical properties has also been displayed. The key features pertaining to both classes have then been unraveled and contrasted. These solutions have been compelled to verify the axisymmetry condition and regularity at the axis and they have been shown to match systematically to exterior vacuum Lewis-Weyl solutions.

Another critical general result has been depicted here in Appendix \ref{omega}. It concerns the interpretation of the $c$ parameter appearing consistently in every solution of the considered issue as an integration parameter of a combination of the field equations involving $G_{00}$ and $G_{03}$. It has been shown that, under some generic conditions, this parameter represents the rotation scalar of the fluid evaluated on the axis of rotation. It has been shown in Appendix \ref{ApB} that such a property can be used to reduce the initial number of parameters of the solutions. 

Since, for both the regular classes of fluid with axial pressure presented here, the $c$ parameter can be expressed as a mere function of $h_0$, i.e., the value of the ratio $h$ at the axis, the significance of this ratio is enhanced while it appears as the only parameter defining each solution in each class.

Now, the interest of the results of the present article should be enhanced by the consideration of those of Paper 3, for a purely azimuthal pressure, and of Paper 4, for a radial pressure and the reader is induced to read them concomitantly. Of course, due to the nonlinearity of the field equations, it is well known that any decomposition of a generic pressure into its principal stress components, i.e., axial, radial and azimuthal, cannot be boldly used in a linear composition of the corresponding solutions. However, for perturbation or numerical treatments, a complete set of solutions with three anisotropic pressure should be meaningful.

Indeed, possible astrophysical applications to extragalactic jet formation for the axially directed pressure case have been proposed in Paper 1. Another promising domain might be that of cosmic topological defect forming superstrings. This would imply that the radial coordinate $r_{\Sigma}$ used to evaluate the length of the radius of the bounding cylinder tends to the zero limit. Since the junction condition is satisfied whatever the value of  $r_{\Sigma}$ - it imposes the radial pressure to vanish, which is realized by construction  everywhere in these spacetimes - there is no contraindication for $r_{\Sigma}$ to take any value whatever small. In that case, a non regular axis, even resulting in an angular deficit or excess, would be no more a reason for ruling out a solution. More classes could therefore be considered for this purpose.

Finally, we need to specify that the relax of the assumption of rigid rotation leads to a more involved but indeed very interesting issue. It is currently work in progress.

\acknowledgments

The author wants to acknowledge the precious help of Renaud Savalle for running the SageMath and Mathematica softwares used for the calculation of the Kretschmann scalars.

\appendix
\section{Rotation scalar parameter} \label{omega}

It is shown, in this Appendix, that interior spacetimes sourced by a rigidly rotating cylinder of fluid - more generally, spacetimes verifying (\ref{sol2}) - and satisfying an equation of the kind
\begin{equation}
\frac{f'}{f} - g(h)\mu' = 0 \label{A1}
\end{equation}
possesses a rotation scalar whose square is equal to $c^2$ on the axis, $c$ being the integration constant appearing in (\ref{sol2}), which is a generic equation for rigid rotation. Equation (\ref{A1}) can be a Bianchi identity, as in the present article where $g(h) = h$ in (\ref{Bianchi3}), or it can proceed otherwise from the field equations as it will be shown in companion Papers 3 and 4.

The general form of $\omega^2$ has indeed been recalled in (\ref{II67}) as \cite{CS20}
\begin{equation}
\omega^2= \frac{c^2}{f^2 \textrm{e}^{\mu}}. \label{A2}
\end{equation}
Now, (\ref{A1}) can be formally integrated as
\begin{equation}
f = \exp\left(\int_{h_0}^h g(u)\frac{\textrm{d}\mu}{\textrm{d}u} \textrm{d}u \right), \label{A3}
\end{equation}
which yields
\begin{equation}
f \stackrel{0}{=} \exp(0) =1. \label{A4}
\end{equation}
Now, considering the point of view of $\mu$, (\ref{A1}) can be formally integrated as
\begin{equation}
\textrm{e}^{\mu} = \exp\left(\int_{h_0}^h \frac{1}{g(u) f(u)} \frac{\textrm{d}f}{\textrm{d}u} \textrm{d}u \right), \label{A5}
\end{equation}
which gives
\begin{equation}
\textrm{e}^{\mu} \stackrel{0}{=} \exp(0) =1. \label{A6}
\end{equation}

Inserting (\ref{A4}) and (\ref{A6}) into (\ref{A2}) one obtains, as desired,
\begin{equation}
\omega^2 \stackrel{0}{=} c^2. \label{A7}
\end{equation}

\section{Example of a fully integrated solution polluted by the ''regularity condition''} \label{ApB}

In the course of going through this issue, a number of mathematically exact solutions of the field equations have been integrated. Most of them were unable to satisfy the entire set of mathematical and physical constraining conditions that are indeed satisfied by the Paper 1 and Class A solutions. However, one among the constraints that have been imposed on the parameters of the solutions can be misleading if improperly employed. Indeed, the ''regularity condition'' that has been discussed at length in the literature \cite{L94,W96,C00,P96}, can pollute a well-behaved solution such as to make it apparently inappropriate. The construction of another interesting class is therefore described here and, submitted to the ''regularity condition'', ends up as being unable to satisfy a single key property. It is given here as an illustration of the care with which one must handle the ''singularity condition'' and also as an example of an analysis performed in a more complex case than that of Class A. Its construction, following the recipe of Sec. \ref{method}, is summarized below.

First, a function $f(h)$ is chosen as
\begin{equation}
f = \frac{c_f}{(1-h)^4}. \label{B1}
\end{equation}
Together with its first and second derivatives with respect to $r$, it is inserted into (\ref{II1}) which yields
\begin{equation}
h'' - \frac{(1-3h)}{h(1-h)}h'^2 + \frac{c^2}{2c_f^2} \frac{h(1-h)^9}{1+h} = 0, \label{B2}
\end{equation}
which can be integrated as
\begin{equation}
r = \epsilon \frac{c_f}{c} \int _{h_0} ^h \frac{\textrm{d}\theta}{\theta(1-\theta)^2 \sqrt{\left[\ln \frac{(1+\theta)^{32}}{\theta} -26 \theta + 8 \theta^2 - 2 \theta^3  +\frac{\theta^4}{4} \right]}}, \label{B3}
\end{equation}
where $\epsilon=\pm 1$ and which implies
\begin{equation}
h' = \epsilon \frac{c}{c_f} h (1-h)^2 \left[\ln \frac{(1+h)^{32}}{h} - 26h + 8 h^2 - 2 h^3 +\frac{h^4}{4} \right]^{\frac{1}{2}}. \label{B4}
\end{equation}
Then, the derivatives of (\ref{B1}) are substituted into (\ref{II10}) such as to obtain
\begin{equation}
\frac{D'}{D} = \frac{h''}{h'} +\frac{(-1 + 2h + 5h^2)}{h(1-h)(1+h)}h', \label{B5}
\end{equation}
that can be integrated as
\begin{equation}
D = c_D \frac{h'}{h(1-h)^3(1+h)}, \label{B6}
\end{equation}
which becomes, after insertion of (\ref{B4}),
\begin{equation}
D = \epsilon \frac{c c_D}{c_f} \frac{1}{(1-h)(1+h)}\left[\ln \frac{(1+h)^{32}}{h}- 26h + 8 h^2 - 2 h^3 + \frac{h^4}{4}  \right]^{\frac{1}{2}}, \label{B7}
\end{equation}
where $c_D$ is an integration constant. 

Now, (\ref{B1}), (\ref{B4}) and (\ref{B7}) are inserted into (\ref{II13}) to obtain
\begin{equation}
k = \frac{c_f}{(1-h)^4} \left[ c_k - \frac{2c c_D}{c_f^2} \int_{h_0}^h \frac{(1-v)^5}{v(1+v)} \textrm{d}v \right], \label{B8}
\end{equation}
where $c_k$ is an integration constant and which can be integrated as
\begin{eqnarray}
k &=& \frac{c_f}{(1-h)^4} \left\{c_k + \frac{2c c_D}{c_f^2}
\left[\ln \frac{h_0(1+h)^{32}}{(1+h_0)^{32} h} - 26(h_-h_0)  \right. \right. \nonumber\\
&+& \left. \left. 8(h^2 - h_0^2) - 2(h^3 - h_0^3) + \frac{(h^4 - h_0^4)}{4} \right] \right\}. \label{B9}
\end{eqnarray}

Then, $l$ is obtained through (\ref{II14}) as
\begin{eqnarray}
l &=& \frac{c^2 c_D^2}{c_f^3} \frac{(1-h)^2}{(1+h)^2} 
\left[\ln \frac{(1+h)^{32}}{h} - 26h + 8h^2 - 2 h^3 + \frac{h^4}{4}  \right] \nonumber\\
&-& \frac{c_f}{(1-h)^4} \large\{ \quad \large\}^2, \label{B10}
\end{eqnarray}
where $\large\{ \quad \large\}$ is short for the expression $\left\{c_k + \frac{2c c_D}{c_f^2}
\left[\ln \frac{h_0(1+h)^{32}}{(1+h_0)^{32} h} - 26(h-h_0) + 8(h^2 - h_0^2) - 2(h^3 - h_0^3) \right. \right.$ $\left. \left. + \frac{(h^4 - h_0^4)}{4}\right] \right\}$.

Now, the Bianchi identity (\ref{Bianchi3}) is used to calculate $\mu(h)$. With (\ref{B1}) inserted, it reads
\begin{equation}
\mu' = \frac{4h'}{h(1-h)}, \label{B11}
\end{equation}
which can be integrated as
\begin{equation}
\textrm{e}^{\mu} = \frac{c_{\mu}h^4}{(1-h)^4}, \label{B12}
\end{equation}
where $c_{\mu}$ is an integration constant.

The energy density $\rho$ is now obtained from (\ref{II19}), which, with the use of (\ref{B4}), (\ref{B7}) and (\ref{B12}), yields
\begin{eqnarray}
\rho &=& \frac{c^2}{\kappa c_f^2c_{\mu}} \frac{(1-h)^7}{h^4(1+h)^2} \left[-4h \ln \frac{(1+h)^{32}}{h} + 4 - 16h  \right. \nonumber\\
&+& \left. 124 h^2 - 32 h^3 - 12 h^4 + 15 h^5 - 4 h^6 \right]. \label{B13}
\end{eqnarray}
The pressure $P_z$ follows as $\rho$ multiplied by $h$.

\subsection{Axisymmetry and regularity conditions}

Then, the axisymmetry condition $l \stackrel{0}{=} 0$ is implemented and gives the following relation between the parameters
\begin{equation}
\ln \frac{(1+h_0)^{32}}{h_0} -26h_0 + 8h_0^2 -2h_0^3 + \frac{h_0^4}{4} = \frac{c_f^4 c_k^2}{c^2 c_D^2} \frac{(1+h_0)^2}{(1-h_0)^6},\label{B14}
\end{equation}
which implies
\begin{equation}
\ln \frac{(1+h_0)^{32}}{h_0} -26h_0 + 8h_0^2 -2h_0^3 + \frac{h_0^4}{4} \geq 0.\label{B15}
\end{equation}
The expression at the left-hand side of (\ref{B15}) is the same as that in the square root of (\ref{B4}) giving $h'$. Hence, (\ref{B15}) implies that $h'$ is a real function, which is consistent for standard astrophysical applications. Moreover, $h_0>0$ and, more generally, $h>0$ is imposed from the definition of the logarithm function appearing in (\ref{B15}) and in a number of expressions displayed above.

Now, the regularity condition (\ref{II23}) enters into play as
\begin{eqnarray}
&&\frac{4c_f c_k^2}{(1-h_0)^5} - \frac{4 c c_D c_k}{c_f}\frac{1-h_0}{h_0(1+h_0)} +  \frac{c^2 c_D^2}{c_f^3}\frac{(1-h_0)^7}{h_0(1+h_0)^3} 
 \nonumber\\
&+& \frac{4 c^2 c_D^2}{c_f^3} \frac{1-h_0}{(1+h_0)^3}\left[\ln \frac{(1+h_0)^{32}}{h_0} - 26h_0 \right. \nonumber \\
&+& \left. 8h_0^2 - 2 h_0^3 + \frac{h_0^4}{4}  \right] = 0, \label{B16}
\end{eqnarray}
where the axisymmetry condition (\ref{B14}) is inserted to obtain
\begin{equation}
\frac{4c_f c_k^2(2+h_0)}{(1-h_0)^5} - \frac{4 c c_D c_k}{c_f} \frac{(1-h_0)}{h_0} + \frac{c^2 c_D^2}{c_f^3}\frac{(1-h_0)^7}{h_0(1+h_0)^2} =0, \label{B17}
\end{equation}
which has two solutions for $c_k$:
\begin{equation}
c_k = \frac{c c_D}{2c_f^2} \frac{(1-h_0)^6}{h_0(1+h_0)} \frac{(1-\eta + h_0)}{(2+h_0)}, \label{B18}
\end{equation}
where $\eta = \pm 1$.

\subsection{Energy conditions} \label{ec}

For standard astrophysical applications, the weak energy condition $\rho>0$ is needed. An analysis of the expression inside the brackets in (\ref{B13}) shows that it is negative or null for any positive value of $h$. Hence $\rho>0$ implies
\begin{equation}
- c_{\mu} (1-h) \geq 0. \label{B19}
\end{equation}
Anticipating the results of the metric signature constraints, one can consider that $c_{\mu}$ is bound to be positive. Therefore, (\ref{B19}) becomes
\begin{equation}
h \geq +1. \label{B20}
\end{equation}
Hence, not only the weak, but also the strong energy condition, $P_z \geq 0$, are satisfied by this solution.

\subsection{Hydrodynamical properties}

These properties are summarized below. They have been calculated using the general expressions displayed in Sec. \ref{hydroI}. Hence, to obtain the nonzero component of the acceleration vector, (\ref{B1}) and derivative, together with (\ref{B4}), are inserted into (\ref{II62}) and yield
\begin{equation}
\dot{V}_1 = 2 \epsilon \frac{c}{c_f} h (1-h)\left[\ln \frac{(1+h)^{32}}{h} - 26h + 8h^2 - 2 h^3 + \frac{h^4}{4}\right]^{\frac{1}{2}}. \label{B18a}
\end{equation}
Its amplitude follows from inserting the same equations plus (\ref{B12}) into (\ref{II65}) such as to obtain
\begin{equation}
\dot{V}^{\alpha}\dot{V}_{\alpha} = \frac{4c^2}{c_f^2 c_{\mu}} \frac{(1-h)^6}{h^2}\left[\ln \frac{(1+h)^{32}}{h} - 26h + 8h^2 - 2 h^3 + \frac{h^4}{4}\right]. \label{B19a}
\end{equation}
Finally, the rotation scalar squared proceeds from substituting (\ref{B1}) and (\ref{B12}) into (\ref{II67}) and reads
\begin{equation}
\omega^2 = \frac{c^2}{c_f^2 c_{\mu}}\frac{(1-h)^{12}}{h^4}. \label{B20a}
\end{equation}

It has been shown, in Appendix A, that the rotation scalar squared must be equal to the square of the integration constant $c$ when evaluated on the axis. This constraint implemented into (\ref{B20a}) yields
\begin{equation}
c_f^2 c_{\mu} = \frac{(1-h_0)^{12}}{h_0^4}. \label{B21a}
\end{equation}

\subsection{The metric signature}

It has already been stressed in Sec. \ref{sign} that, to obtain a proper Lorentzian metric signature, every metric function must be either positive or negative definite. Their sign is thus studied now.

The constraint (\ref{B21a}) issued from the property of the rotation scalar implies obviously $c_{\mu}>0$ which, from (\ref{B12}), yields $\textrm{e}^{\mu}>0$. This forces the three other metric functions to be positive definite.

The $f$ function is indeed so provided the parameter $c_f$ is itself positive.

Now, the expression (\ref{B9}) for $k$ evaluated at the axis gives
\begin{equation}
k \stackrel{0}{=} \frac{c_f c_k}{(1-h_0)^4}. \label{B22a}
\end{equation}
Since $c_f>0$ and $k>0$ is demanded everywhere, in particular at $r=0$, the result is $c_k>0$.

From the axisymmetry condition (\ref{B14}), an expression for $c_k$ can be derived as
\begin{equation}
c_k = \pm \frac{c c_D}{c_f^2} \frac{(1-h_0)^3}{1+h_0}\left[\ln \frac{(1+h_0)^{32}}{h_0} -26h_0 + 8h_0^2 -2h_0^3 + \frac{h_0^4}{4}\right]^{\frac{1}{2}},\label{B23a}
\end{equation}
which can be inserted into the expression (\ref{B9}) for $k$ that becomes
\begin{eqnarray}
&& k = \frac{c c_D}{c_f}\frac{1}{(1-h)^4} \left\{\pm \frac{(1-h_0)^3}{1+h_0}\left[\ln \frac{(1+h_0)^{32}}{h_0} - 26h_0 \right. \right. \nonumber\\ 
&+& \left. \left. 8h_0^2 - 2h_0^3 + \frac{h_0^4}{4} \right]^{\frac{1}{2}} + 2 \left[\ln \frac{h_0(1+h)^{32}}{(1+h_0)^{32}h} - 26(h-h_0) \right. \right. \nonumber\\ 
&+& \left. \left.  8(h^2 - h_0^2) - 2 (h^3 - h_0^3) + \frac{h^4 -h_0^4}{4}  \right]\right\}. \label{B24a}
\end{eqnarray}
It can also be substituted into (\ref{B10}) which becomes
\begin{eqnarray}
&& l = \frac{c^2 c_D^2}{c_f^3} \left\lgroup \frac{(1-h)^2}{(1+h)^2} 
\left[\ln \frac{(1+h)^{32}}{h} - 26h + 8h^2 - 2 h^3 + \frac{h^4}{4}  \right] \right.\nonumber\\
&-& \left. \frac{1}{(1-h)^4} \left\{\pm \frac{(1-h_0)^3}{1+h_0}\left[ \ln \frac{(1+h_0)^{32}}{h_0} - 26h_0 + 8h_0^2 \right. \right. \right. \nonumber\\
&-& \left. \left. \left. 2 h_0^3 + \frac{h_0^4}{4}\right]^{\frac{1}{2}} + 2 \left[\ln \frac{h_0(1+h)^{32}}{(1+h_0)^{32}h} - 26(h-h_0) \right. \right. \right.\nonumber\\ 
&+& \left. \left. \left. 8(h^2 - h_0^2) - 2 (h^3 - h_0^3) + \frac{h^4 -h_0^4}{4}  \right]\right\}^2 \right\rgroup. \label{B25a}
\end{eqnarray}

Noting that the factor $c c_D/c_f$ occurs as such in the expression for $k$ and squared in that for $l$, one can indeed rescale the $\phi$ coordinate from this factor $c c_D/c_f$ which amounts to writing $c c_D/c_f = 1$ to account for the upper limit $2\pi$ of $\phi$ and that can be written as
\begin{equation}
c_f = c c_D. \label{B26a}
\end{equation}
Substituting (\ref{B26a}) into (\ref{B23a}), one obtains
\begin{equation}
c_k = \pm \frac{(1-h_0)^3}{c_f(1+h_0)}\left[\ln \frac{(1+h_0)^{32}}{h_0} -26h_0 + 8h_0^2 -2h_0^3 + \frac{h_0^4}{4}\right]^{\frac{1}{2}}.\label{B27a}
\end{equation}
The previously derived constraints, $h_0>1$ and $c_k>0$, (\ref{B27a}) impose that the sign indeterminacy $\pm$ reduces to the minus sign.

Then, the case $\eta=+1$ is considered. Inserting (\ref{B26a}) and this value for $\eta$ into (\ref{B18}), one obtains
\begin{equation}
c_k = \frac{(1-h_0)^6}{2c_f(1+h_0)(2+h_0)}, \label{B28a}
\end{equation}
which is equalized to (\ref{B27a}), with $\pm$ replaced by $-$, to obtain
\begin{equation}
\frac{(1-h_0)^3}{2(2+h_0)} + \left[\ln \frac{(1+h_0)^{32}}{h_0} -26h_0 + 8h_0^2 -2h_0^3 + \frac{h_0^4}{4}\right]^{\frac{1}{2}} = 0.\label{B29a}
\end{equation}
However, equation (\ref{B29a}) is invalid since, as it is easy to verify, its left-hand side does not vanish whatever the value of $h_0$. The choice $\eta=+1$ is therefore ruled out and $\eta=-1$ will thus be retained in the following. Therefore, (\ref{B18}) becomes
\begin{equation}
c_k = \frac{(1-h_0)^6}{2c_f h_0(1+h_0)}, \label{B30a}
\end{equation}
and (\ref{B27a}) reads
\begin{equation}
c_k = - \frac{(1-h_0)^3}{c_f(1+h_0)}\left[\ln \frac{(1+h_0)^{32}}{h_0} -26h_0 + 8h_0^2 -2h_0^3 + \frac{h_0^4}{4}\right]^{\frac{1}{2}}.\label{B31a}
\end{equation}
Equalizing both expressions for $c_k$, one obtains
\begin{equation}
 \frac{(1-h_0)^3}{2 h_0} +\left[\ln \frac{(1+h_0)^{32}}{h_0} -26h_0 + 8h_0^2 -2h_0^3 + \frac{h_0^4}{4}\right]^{\frac{1}{2}} = 0,\label{B32a}
\end{equation}
whose only solution is $h_0 \sim 3.701866$, which actually fulfills the energy condition $h \geq +1$ derived in Sec. \ref{ec}.

Now, (\ref{B32a}) inserted into (\ref{B24a}) gives
\begin{eqnarray}
k &=& \frac{1}{(1-h)^4}\left\{2\left[\ln \frac{(1+h)^{32}}{h} - 26h + 8h^2 - 2 h^3 + \frac{h^4}{4}  \right] \right. \nonumber\\ 
&-& \left. \frac{(1-h_0)^6}{2h_0^2(1+h_0)}\right\}. \label{B33a}
\end{eqnarray}
The sign of $k$ is that of the expression in curled brackets where $h_0$ is replaced by its numerical value $h_0 \sim 3.701866$. Since a straightforward analysis shows that this expression is positive and never vanishes whatever the positive value of $h$, the metric function $k$ is positive definite for any value of the $r$ coordinate.

Finally, the metric function $l$ can be written as
\begin{eqnarray}
l &=& \frac{1}{c_f} \left\lgroup \frac{(1-h)^2}{(1+h)^2} 
\left[\ln \frac{(1+h)^{32}}{h} - 26h + 8h^2 - 2 h^3 + \frac{h^4}{4}  \right] \right.\nonumber\\
&-& \left. \frac{1}{(1-h)^4} \left\{- \frac{(1-h_0)^6}{2h_0^2(1+h_0)} + 2 \left[\ln \frac{(1+h)^{32}}{h} - 26h \right. \right. \right.\nonumber\\ 
&+& \left. \left. \left. 8h^2 - 2 h^3 + \frac{h^4}{4}  \right]\right\}^2 \right\rgroup. \label{B34a}
\end{eqnarray}
It must be recalled that, from the axisymmetry condition, $l(h_0) = 0$ and, from the regularity condition, $l'(h_0)=0$. Hence $h_0 \sim 3.701866$ is an extremum of the function $l(h)$. However, an analysis of the $l(h)$ function as given by (\ref{B34a}) shows that, for $h \neq h_0$, this function is strictly negative.

The solution described in this Appendix seems therefore to be ruled out owing to an improper signature of the metric. Note however that its initial metric functions, given by (\ref{B1}), (\ref{B9}), (\ref{B10}) and (\ref{B12}) before applying any ''regularity condition'' do not diverge for $h=h_0$. Hence, a blind addition of such a condition is redundant and imposes on the parameters an extra constraint which forces the metric function $l$ to be negative while its initial expression would obviously be positive for well-chosen ranges of the parameters. This stresses the caution needed when making use of this condition.


%
%


%



\hfill

\end{document}